\documentclass[article,onecolumn,preprintnumbers,amsmath,amssymb,floatfix,superscriptaddress]{revtex4}
\usepackage{graphicx}
\usepackage{rotating}
\usepackage{comment}
\usepackage{dcolumn}
\usepackage{makeidx}
\usepackage{color}
\usepackage{amsmath}
\usepackage{amsfonts}
\usepackage{wrapfig}
\usepackage{braket}
\usepackage{bm}
\usepackage{float}
\usepackage{enumitem}

\usepackage{hyperref}
\hypersetup{colorlinks=true, linkcolor=blue, citecolor=blue, urlcolor=blue}

\usepackage[normalem]{ulem}
\usepackage[usenames, dvipsnames]{xcolor}

\begin{document}

\title{
Tuning of magnetic structures topology via single-ion anisotropy and magnetic field \\ in frustrated 2D semiconductors
}

\author{Danila Amoroso}
\affiliation{\footnotesize Consiglio Nazionale delle Ricerche CNR-SPIN, c/o Universit\`a degli Studi “G. D’Annunzio”, I-66100 Chieti, Italy}
\author{Paolo Barone}
\affiliation{\footnotesize Consiglio Nazionale delle Ricerche CNR-SPIN, Area della Ricerca di Tor Vergata, Via del Fosso del Cavaliere 100, I-00133  Rome, Italy}
\author{Silvia Picozzi}
\affiliation{\footnotesize Consiglio Nazionale delle Ricerche CNR-SPIN, c/o Universit\`a degli Studi “G. D’Annunzio”, I-66100 Chieti, Italy}



\begin{abstract}
{\bf 
The effects of competing magnetic interactions in stabilizing different spin configurations are drawing a renewed attention in order to both unveil emerging topological spin textures and to highlight microscopic mechanisms leading to their stabilization. The possible key role of the two-site exchange anisotropy in selecting specific helicity and vorticity of skyrmionic lattices has only recently been proposed. In this work we explore the phase diagram of a frustrated localized magnet on a two-dimensional centrosymmetric triangular lattice, focusing on the interplay between the two-ion anisotropy (TIA) and the single-ion anisotropy (SIA). The effects of an external magnetic field applied perpendicularly to the magnetic layer are also investigated. By means of Monte Carlo simulations, we find a profusion of different spin configurations, going from trivial to high-order Q skyrmionic and meronic lattices. In closer detail, we find that a dominant role is played by the two-ion over the single-ion anisotropy in determining the planar spin texture, whereas the strength and sign of SIA, together with the magnitude of the magnetic field, tune the perpendicular spin components, mostly affecting the polarity (and, in turn, the topology) of the spin-texture. Our analysis confirms the crucial role of anisotropic symmetric exchange in systems with dominant short-range interactions, at the same time predicting a rich variety of complex magnetic textures that may arise from a fine tuning of competing anisotropic mechanisms.
}
\end{abstract}

\maketitle



\section{Introduction}

Competition between different magnetic interactions leading to the condition of {\em frustration}, 
is a key ingredient for the stabilization of non-collinear and non-coplanar magnetic configurations. 
Anisotropic interactions, in particular, play a crucial role in the formation of exotic and topological spin textures, among which the best known are Bloch or Ne\'el type skyrmions and anti-skyrmions with topological charge $|Q|$ of one~\cite{tokura_review,fert_review}; 
nevertheless topological spin textures go beyond conventional skyrmions, from high-$Q$ (anti)skyrmionic~\cite{bertrand_2016,kristian_prb_2017,current_induced_2017} to halved-$Q$ (anti)meronic spin textures and lattices~\cite{PRB2007_merons, batista2015_merons, tokura_merons, exp_merons,pereira_2019,ingrid_review}. 
The different types of isolated two-dimensional (2D) topological spin textures are primarily characterized by the polarity $p$ 
and vorticity $m$, whose product defines their topology with $Q=m\cdot p$. Polarity is associated to the out-of-plane 
magnetization profile when moving from the core of the topological object to its edge (to infinity in a continuous description of isolated object); vorticity, instead, is determined by the in-plane magnetization rotation and is also referred to as the winding number, 
allowing only for integer values~\cite{ingrid_review,jap_review}. 
Therefore, (anti)skyrmions, which are characterized by reversed magnetization directions when comparing the core with its edge, 
corresponding to a spin configuration wrapping a unit sphere, always display unitary values of polarity $p=\pm 1$ and, hence, integer values of $Q$; in 
(anti)merons, instead, the magnetization of the edge and at the core are directed perpendicularly to each other, corresponding 
thus to the wrapping of only half of the sphere and therefore to halved polarity $p=\pm 1/2$, with topological charges that 
are multiples of half-integer values. 
Note that the sign of the topological charge $Q$ changes under time reversal, reversing the $p$ sign.


Conventional skyrmion lattices are usually observed in chiral magnets as a result of the interplay between 
Heisenberg and Dzyaloshinskii-Moriya (DM)~\cite{DM_moriya} exchange interactions~\cite{nagaosa_2009,tokura_report_2012,kristian_prb_2014,tokura_nanotech_2020,stefan_bimerons}; skyrmionic lattices 
with various, or not {\em a priori} determined, topologies can instead occur in geometrically frustrated 
lattices (such as triangular or Kagome) triggered by competing magnetic exchange interactions and assisted by dominant 
non-chiral interactions, such as easy-axis anisotropy~\cite{okubo_prl,maxim_2015,landau_2016,batista_2016,tokura_2019,tokura_current_2014,adv_mater_2Sk}, long-range dipole–dipole~\cite{current_induced_2017,capic_2019,ingrid_2019} and/or Ruderman–Kittel–Kasuya–Yosida (RKKY)~\cite{motome_0,batista_rev_2016} 
interactions, and thermal or quantum fluctuations~\cite{batista_2014_PRX}. In both cases the stabilization of skyrmions 
is often driven by an external magnetic field perpendicular to the magnetic layer (out-of-plane). Formation of meronic textures, instead, are usually driven by in-plane magnetic field and/or easy-plane anisotropy~\cite{tokura_merons,exp_merons,batista2015_merons,ingrid_bimerons}.
The effect of such in-plane magnetic interactions is indeed to penalize the out-of-plane magnetization of skyrmion-like configurations.
This may affect their size and consequently their overlap when arranged on lattices, causing a fractionalization of topological charge and possibly a transition to meronic lattices\cite{batista2015_merons, tokura_merons, exp_merons,pereira_2019}. 

Spontaneous stabilization of skyrmionic lattices have been proposed so far only in itinerant magnets, displaying long-range exchange interactions mediated by conduction electrons~\cite{robler_2006,batista_2008,stefan_2011,motome_1}. 
Only recently, a frustrated semiconducting 2D-magnet, NiI$_2$, has been predicted to display the spontaneous formation of a stable high-$Q$ skyrmionic lattice ($Q=2$)~\cite{noi}, with well defined topology and chirality determined by the anisotropic part of the short-range symmetric exchange, in absence of DM and Zeeman interactions.
In particular, authors pointed out that the anisotropic symmetric exchange, also referred to as two-ion anisotropy or bond-dependent exchange anisotropy~\cite{motome_4}, may act, in the presence of magnetic frustration, as an emergent chiral interaction when displaying noncoplanar principal axes (Fig.~\ref{profile_ground}d), thus adding frustration in the relative orientation of spins and determining the topology of the localized spin texture.\\

Given these premises, in this work we elaborate on our recent findings by investigating the effects of the interplay between the two-ion (TIA) and single-ion (SIA) anisotropies on the formation and stabilization of topological spin structures. A rich phase diagram is uncovered by means of Monte Carlo simulations, where both SIA and applied out-of-plane magnetic field trigger diverse transitions between different kinds of magnetic configurations with the topological charge ranging from high-$Q$ to zero values. The spontaneous high-$Q$ skyrmion lattice previously identified is found to extend over a significant ranges of SIA, confirming the robustness of the TIA-based mechanism: on the one hand, the in-plane component of all non-coplanar emerging triple-$\bm q$ spin states is found to be always fixed by the two-ion anisotropy, confirming its primary role in determining the rotational direction of in-plane magnetization over the SIA; on the other hand, the out-of-plane magnetization is tuned by both SIA and applied magnetic field, thus affecting mostly the polarity and, hence, the spin-textures topology. It is worth to note that similar magnetic phases have been recently reported also in frustrated itinerant magnets~\cite{motome_phases}, arising from the interplay of magnetic anisotropies and dominant long-range  and higher-order spin interactions, such as biquadratic exchange, mediated by conduction electrons and spin-charge coupling, as opposed to the short-range interaction between local magnetic moments considered here.

\section{Materials and Methods}

The starting point of our analysis is the classical spin Hamiltonian~(given in the following Eq.~\ref{hamiltonian1}) describing the magnetic properties of the semiconducting and centrosymmetric NiI$_2$ monolayer~\cite{noi}. NiI$_2$ belongs to the family of transition-metal-based ({\it M}) van der Waals materials~\cite{huang_2017,mc_guire,2D_hot,dm_cri3,adv_mater_2d,han_gct,ni2_exp1,ni2_exp2}, comprising layers where magnetic cations form a 2D triangular lattice and thus displaying {\em geometrical frustration}~\cite{frustration_1,frustration_2}. Our previous first-principles based investigation on NiI$_2$ monolayer revealed that its magnetic properties are ruled by:

\begin{enumerate}[label=(\roman*)]
\item strong magnetic frustration from competing nearest-neighbor ($J^{1iso}$) ferromagnetic (FM) and third-nearest neighbor ($J^{3iso}$) anti-ferromagnetic (AFM) exchange interactions; 

\item very weak easy-plane single-ion anisotropy against highly anisotropic symmetric exchange or {\it two-ion anisotropy}.
\end{enumerate}

Monte Carlo (MC) simulations have showed then that the combination of these two properties results in the spontaneous stabilization of a topological spin-structure with well-defined topology and chirality.\\

More in detail, the model Hamiltonian for classical spins of unit length $s_i=1$ is defined as:
\begin{equation}
	H=\frac{1}{2}\sum_{i\neq j}\mathbf{s}_i \mathbf{J}_{ij} \mathbf{s}_j+\mathbf{s}_i\mathbf{A}_{i}\mathbf{s}_i
	\label{hamiltonian1}
\end{equation} 
where {\bf J}$_{ij}$ and {\bf A}$_i$ are tensors describing the exchange interaction and single-ion anisotropy, respectively~\cite{mike_2013}. It is convenient to decompose the exchange coupling tensor into an isotropic part $J^{iso}_{ij}=\frac13\mathrm{Tr}\mathbf{J}_{ij}$ and an anisotropic symmetric part $\mathbf{J}^{S}_{ij}=\frac12(\mathbf{J}_{ij}+\mathbf{J}^{\mathrm{T}}_{ij})-J^{iso}_{ij}\mathbf{I}$, herein also referred to as two-ion anisotropy (TIA). Due to the  the inversion symmetry of the lattice (with $D_{3d}$ point group), the DM interaction, corresponding to the antisymmetric exchange term $\mathbf{J}^A_{ij}=\frac12(\mathbf{J}_{ij}-\mathbf{J}^{\mathrm{T}}_{ij})$, is identically zero.
Reference magnetic parameters for NiI$_2$ monolayer evaluated from density-functional theory (DFT) calculations~\cite{noi}, are given in Table~\ref{exchange_param}. 
Technical details about DFT calculations and the estimate of magnetic parameters can be found in the Methods section of Ref.\cite{noi}.

\begin{table}[t]
\centering
\vspace{.5cm}
\begin{tabular}{c|cccccc|c|c}
\centering
 $J^{1iso}$ & $J^{S}_{xx}$  & $J^{S}_{yy}$  &  $J^{S}_{zz}$  & $J^{S}_{yz}$ & $J^{S}_{xz}$  &  $J^{S}_{xy}$ & $J^{3iso}$ & $A_{zz}$ \\[2pt]
  -7.0 & -1.0 & +1.4 & -0.3 & -1.4 & 0.0 & 0.0 & +5.8 & +0.6 \\
 \botrule
\end{tabular}
\caption{\small 
 Magnetic exchange coupling parameters and single-ion anisotropy, in term of energy unit (meV), for NiI$_2$ monolayers. Interactions have been estimated by means of DFT calculations based on the four-state energy mapping method~\cite{mike_2013}, within the PBE+U approximation~\cite{pbe,liet_U} ($U$=1.8~eV, $J$=0.8~eV), via the VASP code~\cite{kresse_vasp, vasp_site}. The matrix elements of the two-ion exchange anisotropy~($\mathbf{J}^{S}$ or TIA) between nearest-neighbour spins, are expressed in a local
cartesian $\{x, y, z\}$ basis, where $x$ is parallel to the Ni-Ni bonding vector (see Fig.~2a in Ref.~\cite{noi} and Fig.~S1 in Ref.~\cite{noi_suppl}).
In the adopted convention, negative (positive) values of the exchange parameters refer to FM (AFM) magnetic interaction, while positive (negative) value of SIA ($A_{zz}$) indicates easy-plane (easy-axis) anisotropy. }
	\label{exchange_param}
\end{table}

Taking into account such exchange interactions, the phase diagram of the magnetic system arising when tuning the single-ion anisotropy and applied out-of-plane magnetic field has been studied within a Monte Carlo approach. In particular, MC calculations were performed using a standard Metropolis algorithm on $L\times L$ triangular supercells with periodic boundary conditions. 
Starting from high temperature ($T$), at each simulated $T$, we used 10$^5$ MC steps for thermalization and 5$\times$10$^5$ MC steps for statistical averaging. 
The lateral size of the simulation supercells has been chosen as $L = n L_{m.u.c.}$, where $n$ is an integer and $L_{m.u.c.}$ is the lateral size, in units of the lattice constant $a_0$, of the magnetic unit cell needed to accommodate the lowest-energy non-collinear helimagnetic spin configurations. Accordingly, we estimated $L_{m.u.c.}$ as $1/q$, where $q$ is the length of the propagation vector {\bf q} minimizing the isotropic exchange interaction in momentum  space $J(\mathbf{q})$. 
Using the magnetic parameters listed in Table~\ref{exchange_param}, the propagation vector for the isotropic model is given by $q = 2 \cos^{-1}[(1+\sqrt{1-2J^{1iso}/J^{3iso}})/4]$~\cite{batista_rev_2016,batista_2016}, resulting in $L_{m.u.c.}\simeq 8$. Presented results are thus obtained by calculations performed on supercells with lateral size $L=3L_{m.u.c.}=24$. 

Further insight on the magnetic configurations is obtained by evaluating the spin structure factor:
\begin{eqnarray}
S(\mathbf{q}) &=& \frac{1}{N}\,\sum_{\alpha=x,y,z}\,\left\langle \left\vert \sum_i s_{i,\alpha}\,e^{-i\mathbf{q} \cdot \mathbf{r}_i}\right\vert^2\right\rangle
\end{eqnarray}
where $\mathbf{r}_i$ denotes the position of spin $\mathbf{s}_i$ and $N=L^2$ is the total number of spins in the supercell used for MC simulations. The bracket notation is used to denote the statistical average over the MC configurations. The spin structure factor provides direct information on the direction and size of the propagation vectors. 
The topological nature of the multiple-$q$ phase has been assessed by evaluating the topological charge (skyrmion number) of the lattice spin field of each supercell as $ \langle Q \rangle = \langle \sum_i \Omega_i\rangle$, where $\Omega_i$ is calculated for each triangular plaquette as~\cite{discrete_topological}:

\begin{eqnarray}
\tan\left(\frac{1}{2} \Omega_i \right)=\frac{\mathbf{s}_1\cdot \mathbf{s}_2 \times\mathbf{s}_3}{1+\mathbf{s}_1\cdot \mathbf{s}_2+\mathbf{s}_1\cdot \mathbf{s}_3+\mathbf{s}_2\cdot \mathbf{s}_3}
\end{eqnarray}


In the following, starting from the exchange interactions estimated in the prototypical monolayer NiI$_2$ as representative example of highly frustrated semiconducting 2D triangular lattice systems, we explore the magnetic phase diagram as a function of the strength and direction of the SIA as well as of an applied out-of-plane magnetic field $B_z$.

\section{Results} 

\begin{figure*}[th!]
\centering
\includegraphics[width=15cm]{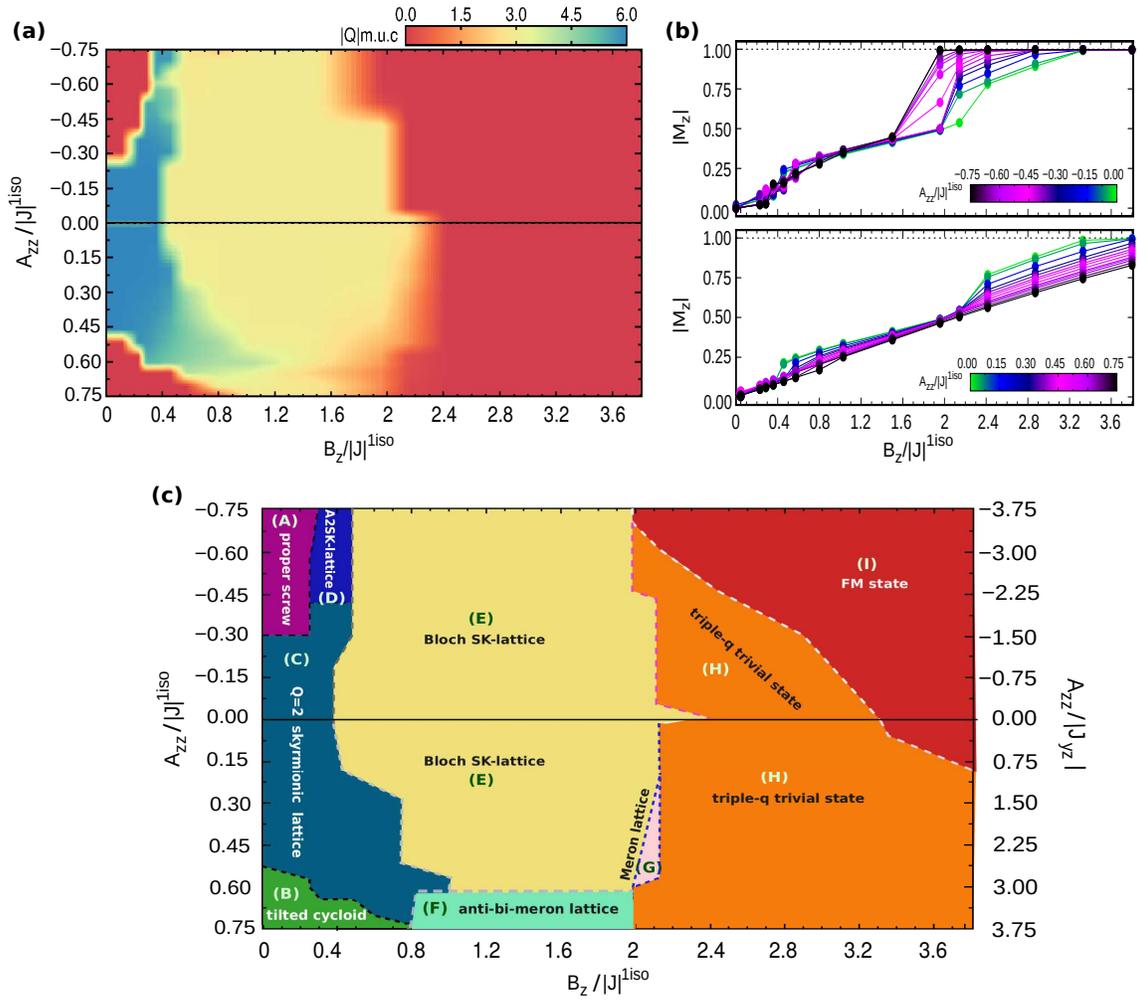}
\caption{Single-ion anisotropy and field induced phase transitions. {\bf (a)} Low temperature ($T=1$~K) phase diagram in the SIA-magnetic field $A_{zz}-B_z$ parameters space (both $A_{zz}$ and $B_z$ expressed in units of the nearest neighbour isotropic exchange interaction $\vert J\vert^{1iso}$). Color-map indicates the absolute value of total topological charge densities per magnetic unit cell $|Q|_{m.u.c.}$. 
{\bf (b)} Evolution of out-of-plane magnetization $M_z$ as a function of the magnetic field $B_z$ for increasing SIA, easy-axis ($A_{zz}/|J|^{1iso}<0$, top) and easy-plane ($A_{zz}/|J|^{1iso}>0$, bottom) respectively.
{\bf (c)} Schematic $A_{zz}-B_z$ phase diagram identifying regions occupied by the different spin configurations, labelled with capital letters. Phase boundaries must be regarded only as semi-quantitative, as field-induced topological phase transitions go through almost continuous transformations of the spin-texture and metastable spin states with fractionalized topological charge, that prevent the identification of accurate and sharp phase boundaries. On the right-side of the $y$-axis $A_{zz}/|J_{yz}|$ ratio is also reported. The horizontal solid line separates the easy-axis (above) and easy-plane (below) SIA regions.  
\label{phase_diagram}}
\end{figure*} 


In Fig.~\ref{phase_diagram}, we show the low temperature ($T=1$~K) magnetic phase diagram in the SIA-field ($A_{zz}\textrm{-}B_z$) plane, obtained by fixing the exchange coupling interactions to the reference values reported in Table~\ref{exchange_param} and varying value and direction of the single-ion anisotropy; for each $A_{zz}$ value, increasing magnetic field ($B_z$) was also applied along the perpendicular direction of the monolayer triangular lattice. As shown in Fig.~\ref{phase_diagram}(a), for each combination of exchange coupling and $A_{zz}$ parameters there are field-induced topological phase transitions between states with different topological charge $Q$, listed below, approximately, per magnetic unit cell (m.u.c): 

\begin{table}[h]
\centering
\begin{tabular}{c|l}
 $-0.75  \lesssim A_{zz}/|J|^{1iso}  \lesssim  -0.30$ & $Q_{m.u.c}=0\to6\to3\to0$ \\
 $-0.30 \lesssim A_{zz}/|J|^{1iso} \lesssim 0.30$ & $Q_{m.u.c}=6\to3\to0$ \\
  $0.30 \lesssim A_{zz}/|J|^{1iso} \lesssim 0.50$ & $Q_{m.u.c}=6\to3\to1.5\to0$ \\
 $0.50 \lesssim A_{zz}/|J|^{1iso} \lesssim 0.65$ & $Q_{m.u.c}=0\to6\to3\to0$ \\  
 $0.65 \lesssim A_{zz}/|J|^{1iso} \lesssim 0.75$ & $Q_{m.u.c}=0\to3\to0$ \\
\end{tabular}
\end{table}

As schematically represented in Fig.~\ref{phase_diagram}(c), the phase diagram consists of various trivial and topologically equivalent phases - {\it i.e.} displaying the same $Q_{m.u.c}$ - but characterized by different kind of spin configurations. As shown in Fig.~\ref{phase_diagram}(b), for different realizations of easy-axis and easy-plane anisotropy, field-induced phase transitions are accompanied by discontinuities of the out-of-plane magnetization $M_z$.
Within the considered range of applied $B_z$, the magnetization saturation $M_z=1$, corresponding to all spins aligned parallel to the magnetic field, is reached only in the case of easy-axis anisotropy and for small easy-plane anisotropy; the latter penalizes in fact the out-of-plane component of spins. In-plane magnetization is zero in all cases. \\

In the following we describe and analyze in details the various anisotropy-field induced magnetic orders. 

\begin{figure}[t]
\centering
\includegraphics[width=14cm]{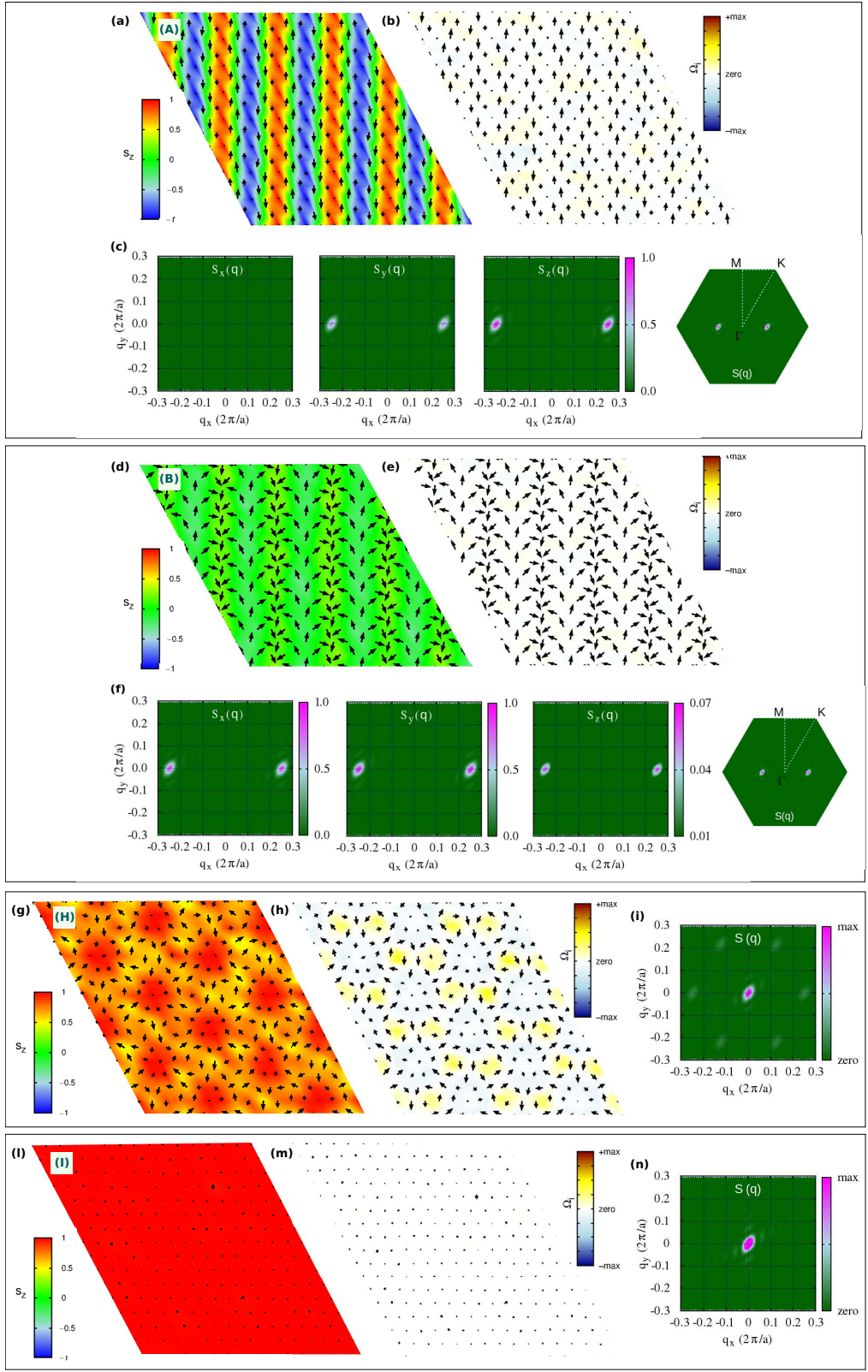}
\caption{ Trivial spin states with $Q_{m.u.c}=0$. Snapshots at $T=1$~K from Monte-Carlo simulations of real-space spin configurations and corresponding topological charge densities $\Omega_i$, in sequence: black arrows represent in-plane $\{s_x,s_y\}$ components of spins; colormap indicates the out-of-plane $s_z$ spins component in the first snapshot and $\Omega_i$ in the second one. The associated spin structure factor $S$({\bf q}) is also shown; its decomposition in the Cartesian components $S_x(\mathbf{q}),~S_y(\mathbf{q})$ and $S_z(\mathbf{q})$ is shown for the spin spirals configurations (first and second panels). Capital letters on each spin-configuration refer to the labels used in the phase diagram Fig.~\ref{phase_diagram}c to name each spin configuration: (A)-phase single-({\bf q}) spiral (proper screw), (B)-phase single-({\bf q}) spiral (tilted cycloid), (H)-phase triple-({\bf q}) trivial state with FM component at $\Gamma$-({\bf q}$=0$), (I)-phase FM state. \label{zeroQ}}
\end{figure}

\subsection{Topologically trivial spin orderings.}

We start our analysis by discussing the topologically trivial spin configurations found in the phase diagram in the strong-field region and at weak fields but for large SIA (both easy-axis and easy-plane).
In Fig.~\ref{zeroQ} we show both spin configurations and topological charge densities $\Omega_i$ of such phases, labeled as (A), (B) and (H) in the phase diagram of Fig.~\ref{phase_diagram}(c), alongside the associated spin structure factor. \\

At weak fields, two helical single-$\bm q$ states appear for strong easy-axis (A) and easy-plane (B) SIA, whose spin configurations are shown in
Fig.~\ref{zeroQ}(a) and Fig.~\ref{zeroQ}(d). No topological charge appears in such phases, as clearly shown in Figs. \ref{zeroQ}(b) and (e). From Figs. \ref{zeroQ}(c) and (f), the propagation vector is $\bm q_1=(-2\delta,\delta)$, with $q/2=\delta \simeq 1/8$ in reduced coordinates, being mostly determined by frustrated isotropic exchange. Due to the three-fold rotational symmetry of the triangular lattice, the single-$\bm q$ helices propagating along the $x$-axis with $\bm q_1$ are equivalent, and hence energetically degenerate, to helices propagating along the symmetry-equivalent directions rotated by $+ 120^{\circ}$, with $\bm q_2=(\delta, -2\delta)$, and $-120^{\circ}$, with  $\bm q_3=(\delta, \delta)$. The plane of spins rotation is selected by both SIA and TIA, which thus determine the nature of the helical states. Inspection of both spin structures and spin structure factors reveals that: phase (A) consists of a proper-screw spiral propagating along the Cartesian $x$-axis, {\it i.e.} along the {\it M}-{\it M} bond direction, with spins rotating in the perpendicular $yz$-plane; phase (B) is a tilted cycloid, {\it i.e.} a helix where spins rotate in a plane containing the propagation vector (parallel to $x$) but tilted around it, causing the spins to acquire a non-zero $z$-component highlighted by the peaks of $S_z(\bm q)$ shown in Fig.~\ref{zeroQ}(f). 
Modeling of the single-$\bm q$ proper-screw and cycloidal helices is presented in Appendix~\ref{model_spiral}.

Combination of the three spirals defines a triple-$\bm q$ state, that is the case for the (H)-phase depicted in Fig.~\ref{zeroQ}g. This is a trivial non-collinear spin configuration, but characterized by local nonzero scalar chirality as evident from the colormap of the topological charge density in Fig.~\ref{zeroQ}h; it is possible to recognize, in fact, a hexagonal lattice formed by six vortices ($m$=+1) with chirality opposite to the surrounded anti-bi-vortex core ($m$=-2). This state occupies a wide region of the phase diagram for high values of applied magnetic field, before 
evolving into a pure ferromagnetic state [phase (I), Fig.~\ref{zeroQ}l]. In addition to the weak peaks at $\bm q_1$, $\bm q_2$, and $\bm q_3$ (Fig.~\ref{zeroQ}n), indeed this (H)-phase displays an out-of-plane ferromagnetic component as from the relevant peak in the $S(\bm q)$ at the $\Gamma$-point ($\bm q=0$), which is also reflected in the remarkable out-of-plane component of the magnetization, $0.5 \lesssim M_z < 1$ (Fig.~\ref{phase_diagram}b), and in the $s_z$ profile depicted in Fig.~\ref{profile}o.
Noteworthy, such triple-$\bm q$ trivial state is largely favored by the easy-plane anisotropy, which indeed competes with $B_z$ penalizing the out-of-plane spin component; for $A_{zz}/|J|^{1iso}>0.15$ ($A_{zz}/|J_{yz}|>0.75$), the FM state is in fact never achieved within the wide range of explored $B_z$. Conversely, the easy-axis anisotropy sustains the $B_z$ action, favoring the out-of-plane spin orientation; in fact, the region of the phase diagram occupied by the (H)-phase reduces until disappearing for $A_{zz}/|J|^{1iso}\lesssim -0.75$ (Fig.~\ref{phase_diagram}c).

\begin{figure}[t]
\centering
\includegraphics[width=14cm]{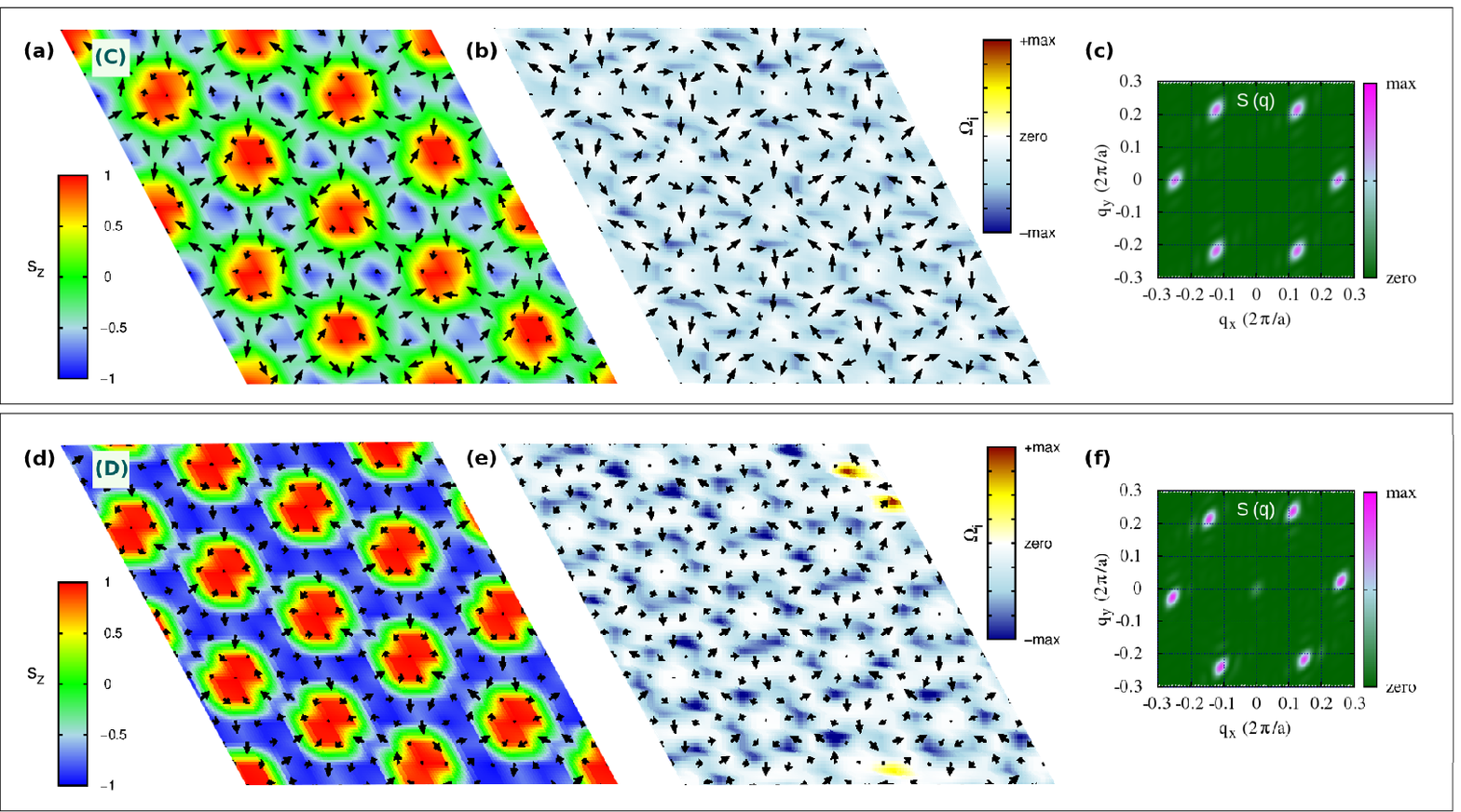}
\caption{Topological lattices with $|Q|_{m.u.c}=6$. Snapshots at $T=1$~K from Monte-Carlo simulations of real-space spin configurations and corresponding topological charge densities $\Omega_i$, in sequence: black arrows represent in-plane $\{s_x,s_y\}$ components of spins; colormap indicates the out-of-plane $s_z$ spins component in the first snapshot and $\Omega_i$ in the second one. The associated spin structure factor $S$({\bf q}) is also shown. Capital letters on each spin-configuration refer to the labels used in the phase diagram Fig.~\ref{phase_diagram}c to name each spin configuration: (C)-phase, triple-({\bf q}) state, composed by periodic repetition of two vortices (V, $m=+1$) and an anti-bi-vortex (A2V, $m=-2$) with total $Q=2$; (D)-phase, triple-({\bf q}) state, consisting of a quasi-ideal anti-bi-skyrmions (A2SK) lattice. 
\label{seiQ}}
\vspace{0.5cm}
\includegraphics[width=14cm]{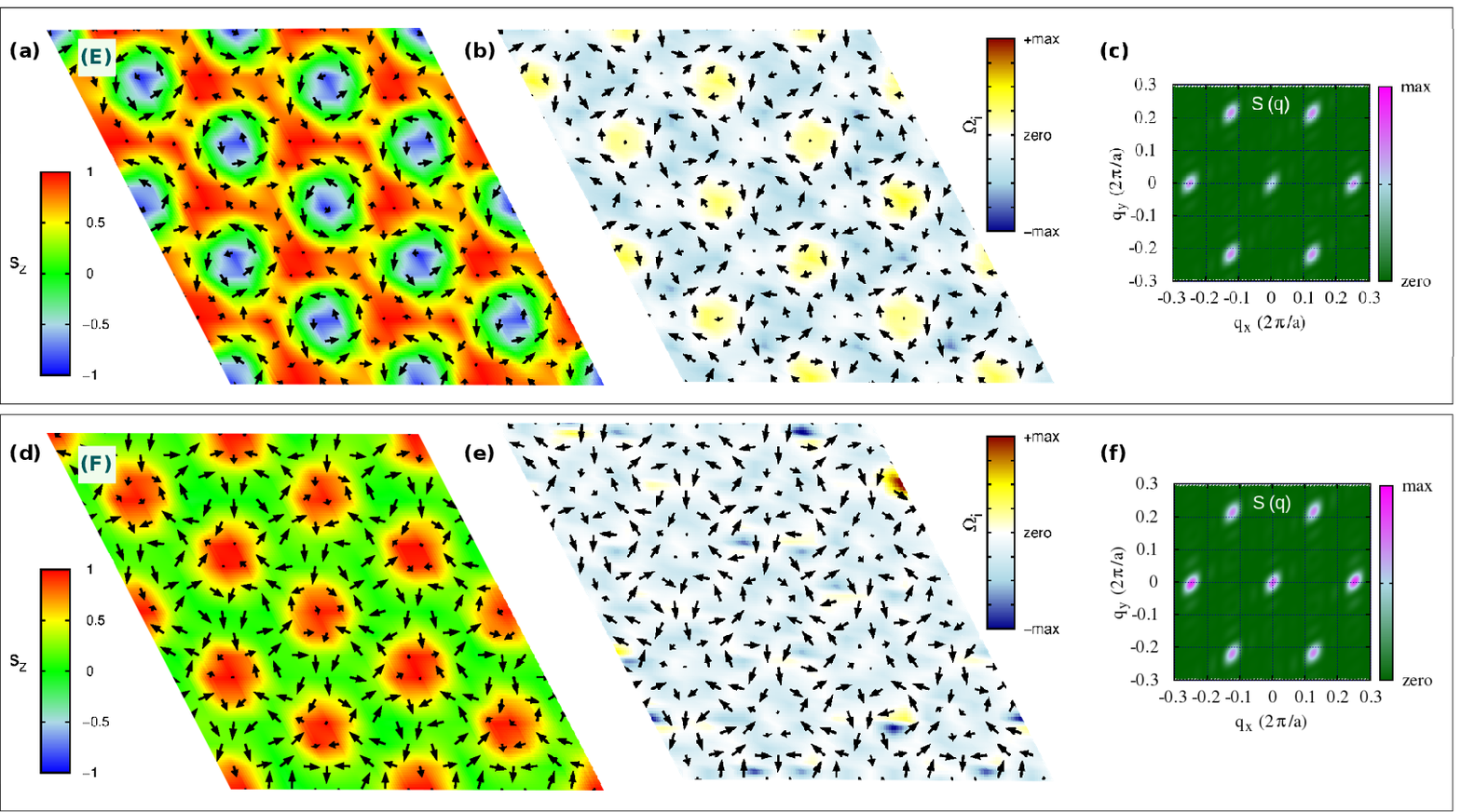}
\caption{Topological lattices with $|Q|_{m.u.c}=3$. Plots as in Fig.~\ref{seiQ} for the field-induced triple-({\bf q}) states: (E)-phase, consisting of a Bloch-type skyrmions (SK) lattice; (F)-phase consisting of a anti-bi-merons (A2M) lattice. The peak at ${\bm q}=0$ in the $S$({\bf q}) reflects the ferromagnetic component induced by the applied $B_z$.  \label{treQ}} 
\end{figure}
\begin{figure}[]
\centering
\includegraphics[width=14cm]{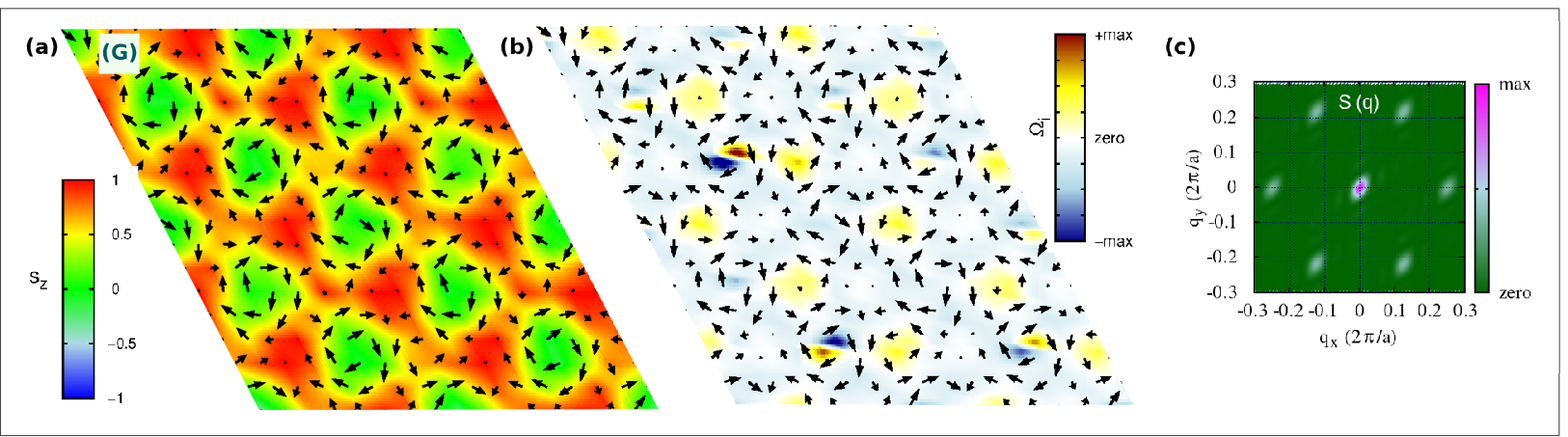}
\caption{Topological lattice with $|Q|_{m.u.c}=1.5$. Plots as in Fig.~\ref{seiQ} for the (G)-phase, triple-({\bf q}) state, consisting of a lattice of merons from the action of large easy-plane anisotropy and $B_z$, as from the ${\bm q}=0$-peak in the $S$({\bf q}), on the SK-lattice of the (E)-phase.  \label{mezzoQ}} 
\end{figure}

\subsection{Topological spin orderings.}

In this section we turn our attention to the diverse topological spin configurations which can be realized by tuning the competition between the single-ion anisotropy and the applied field, whose spin-textures, topological charge densities and spin structure factors are shown in Fig.~\ref{seiQ}, Fig.~\ref{treQ} and Fig.~\ref{mezzoQ}. All topological lattices are 
triple-{\bf q} states, with {\bf q}$_1=(-2\delta,\delta)$, {\bf q}$_2=(\delta,-2\delta)$, {\bf q}$_3=(\delta,\delta)$ and $\delta\simeq 1/8$ [Fig.~\ref{seiQ}(c,f), Fig.~\ref{treQ}(c,f), Fig.~\ref{mezzoQ}(c)],
characterized by atomic scale skyrmionic or meronic structures composed by nano-sized topological objects with a radius counting $\simeq 4$ spins, and thus a diameter of about few units of the lattice parameter $a_0$, as determined by the strength of the magnetic frustration ({\it i.e.} $J^{3iso}~vs~J^{1iso}$ as discussed in the previous Methods section).

\subsubsection{Topological lattice with $Q_{m.u.c}=6$.}
In Fig.~\ref{seiQ}a and Fig.~\ref{seiQ}d we show the high-$Q$ topological lattices labelled phases (C) and (D) respectively, which occupy the region of the phase diagram for zero and low magnetic field; in particular, the (C)-phase is the spontaneous topological phase reported in \cite{noi} stabilized by the exchange magnetic couplings given in Table~\ref{exchange_param}. \\

In closer detail, the topological lattice of the (C)-phase ranges over $-0.30\lesssim A_{zz}/|J|^{1iso}\lesssim +0.50$ and $-1.50\lesssim A_{zz}/|J_{yz}|\lesssim +2.50$ for zero field, whereas it slightly extends over the easy-plane anisotropy part of the phase diagram upon low $B_z$, as shown in Fig.~\ref{phase_diagram}(a,c). It is a triple-{\bf q} state characterized by a hexagonal lattice formed by six vortices (V, $m$=+1) with downward central spin, hosting at the center of each hexagon an anti-bi-vortex (A2V, $m$=-2) with upward core, as shown in Fig.~\ref{seiQ}a. Such spin texture defines an homogeneous topological charge density, as depicted in Fig.~\ref{seiQ}b, hence uniform scalar chirality, and gives rise to a topological charge of six per magnetic unit cell. This topological lattice can be regarded as the periodic repetition of a composition of topological objects consisting of two vortices and one anti-bi-vortex (schematically depicted in Fig.~\ref{profile_ground}g), all contributing to defining a global $Q=2$. 
As evident from the $s_z$ profile reported in Fig.~\ref{profile_ground}(a), the central A2V does not display in fact an uniform, unitary polarity along its surrounding perimeter (highlighted as a purple circle traced in Fig.~\ref{profile_ground}b): the upward spin of the core ($s_z=+1$) is not reversed downward ($s_z=-1$) at all points of its finite edge. Therefore, it brings a fractionalized topological charge $Q_{A2V}=-2\cdot p_{A2V}$, with $0<p_{A2V}<+1$. The missing $Q$ fraction is carried by the two vortices, which, similarly, bring a fractionalized charge $Q_{V}=+1\cdot p_{V}$, with $-1<p_{V}<0$. Accordingly, $Q=Q_{A2V}+2Q_{V}=2$.
The minimal magnetic cell accommodating this topological lattice consists of three repeated objects compositions; hence $|Q|_{m.u.c}=6$. The topological lattice can be therefore interpreted as a fractionalized anti-bi-skyrmion (A2SK) lattice, where the fractionalization of the topological charge can be ascribed to their close packing leading to the incomplete spin wrapping highlighted in Fig.~\ref{profile_ground}~\cite{batista2015_merons}. A similar realization of a fractionalized skyrmion lattice has been recently reported in MnSc$_2$S$_4$, where an applied field has been experimentally shown to stabilize a lattice of fractionalized Bloch-type skyrmions and incipient merons~\cite{gao_nature2020}.

By increasing the easy-axis anisotropy, {\it i.e.} for $-0.75\lesssim A_{zz}/|J|^{1iso}\lesssim -0.45$ and $-3.75\lesssim A_{zz}/|J_{yz}|\lesssim -2.25$, but keeping low values of the applied magnetic field, the magnetic phase transforms into the (D)-phase shown  in Fig.~\ref{seiQ}d. Despite the weak FM component seen in the spin structure factor (Fig.~\ref{seiQ}f), compatible with both applied field and enhanced easy-axis anisotropy, the spin configuration approaches an ideal triangular lattice of anti-bi-skyrmions: each anti-bi-vortex is surrounded by a magnetic background where spins are fully reversed with respect to the core of the A2V along almost all radial directions, as depicted in Fig. \ref{profile}(a,b). Nevertheless, the weak A2SKs overlap due to the small size of the topological objects, namely their proximity, still produces residual vortices at the center of the triangles formed by three nearest-neighbor A2SK, and hence a small fractionalization of the topological charge~\cite{batista2015_merons}, with $1.5\lesssim|Q_{A2SK}|<2$ (or equivalently $4.5\lesssim|Q|_{m.u.c}<6$ considering three anti-bi-skyrmions per magnetic unit cell). 
Accordingly, the topological charge density map is no longer homogeneous, but the highest intensities are localized around the anti-bi-vortex core (Fig.~\ref{seiQ}e). 
We finally notice that the cores of each A2SK is found to align anti-parallel to the applied field direction. The spin texture shown in Fig.~\ref{seiQ}d has been obtained by applying a negative $B_z$ to ease the comparison with the (C)-phase displayed in Fig.~\ref{seiQ}a; we verified that reversal of the magnetic field systematically causes a reversal of the A2SK-core magnetization and an alignment of the magnetic background with the field. 


\subsubsection{Topological lattice with $Q_{m.u.c}=3$.}

In Fig.~\ref{treQ}a and Fig.~\ref{treQ}d we show the topologically equivalent $|Q|_{m.u.c}=3$ lattices, labelled phases (E) and (F) respectively, that occupy a wide portion of the phase diagram in an intermediate range of applied magnetic field $B_z$ (Fig.~\ref{phase_diagram}c). Despite they carry the same topological charge, their spin textures are markedly different. \\

As shown in Fig.~\ref{treQ}a, the (E)-phase consists of a triangular skyrmion lattice, with vortices and anti-bi-vortices appearing at the intersection of clearly visible Bloch-like skyrmions, namely, at the center of the triangles that build up the skyrmion lattice. Such phase can be seen as a direct transition of the stable (C)-phase induced by intermediate values of the applied perpendicular field: upon $B_z$, spins of the anti-bi-vortex core remains parallel to the field, spins of one of the two vortices also tend to align with the magnetic field, while spins at the core of the other vortex remain anti-parallel. Such a change in the $s_z$ spins component in one of the two vortices affects the spin scalar chirality, leading thus to the topological transition from $|Q|_{m.u.c}=6$ in the (C)-phase to $|Q|_{m.u.c}=3$ in this (E)-phase. In the latter phase, the topological charge density map shows in fact negligible contribution from the A2V, while opposite sign of $\Omega_i$ for the upward vortex and the downward one. The highest contribution to the topological charge arises from the vortices whose core-magnetization remains antiparallel to the applied field, thus defining the Bloch-type skyrmion lattice. Indeed, the corresponding $s_z$ profile shown in Fig.~\ref{profile}(g,h) testifies for an almost full reversal of the out-of-plane spin component, {\it i.e.} polar angle variation of about $180^{\circ}$, when moving from the core ($s_z=-1$) to the edge ($s_z=+1$) of the Bloch-type vortex.


Such skyrmion lattice is found to be robust with respect to easy-axis anisotropy, whereas, a strong enough single-ion easy-plane anisotropy ($A_{zz}>>0$) induces its transformation into the (F)-phase, observed in a narrow region of the investigated phase diagram (see Fig.~\ref{phase_diagram}c): the out-of-plane component of magnetization antiparallel to the applied field is here further suppressed by the strong easy-plane SIA. In particular, the analysis of the spin texture shown in Fig.~\ref{treQ}d, combined with the $s_z$ profile (Fig.~\ref{profile}d,e) and the topological charge, indicates that (F)-phase is a lattice of anti-bi-merons (A2M): the A2V of the (C)-phase preserves the upward core, with spins parallel to $B_z$, while the surrounding vortices preserve only the planar spins components; indeed the polar angle changes by $90^{\circ}$ when moving along the radial direction from the core to the edge, {\it i.e.} $p=+0.5$, as from the $s_z$ profile in Fig.~\ref{profile}d. As a consequence, the topological charge of the (F)-phase is halved with respect to that of the (C)-phase, that is $|Q|_{m.u.c}=6\to3$. To the best of our knowledge, this is the first time that such kind of A2M texture is reported.

\subsubsection{Topological lattice with $Q_{m.u.c}=1.5$.} 

As the last relevant topological phase composing the phase diagram, we show in Fig.~\ref{mezzoQ}a a meronic lattice with $Q_{m.u.c}=1.5$, which can be regarded as an intermediate spin configuration during the evolution of the SK-lattice of the (E)-phase into the trivial triple-({\bf q}) ferromagnetic (H)-phase upon $B_z$ (Fig.~\ref{phase_diagram}c). The competition of the intermediate-strong easy-plane anisotropy with the intermediate-strong applied field causes the transformation of the Bloch-like skyrmion into a meron with halved topological charge: as the spin texture in Fig.~\ref{mezzoQ}a and the $s_z$ profile of the vortex core in Fig.~\ref{profile}(l,m) show, the latter loses the spin component along the $z$-direction, while it remains aligned to the field in the surrounding vortices and anti-bi-vortices, which occupy alternatively the center of the triangles formed by the merons: spins at the edge of the topological vortex are directed perpendicularly with respect to its core, halving the topological charge of the Bloch-like skyrmion lattice of the (E)-phase, $|Q|_{m.u.c}=3\to1.5$. 

\subsection{Field-induced phase transition from two-ion anisotropy tuning. } 

\begin{figure}[b]
\centering
\includegraphics[width=12cm]{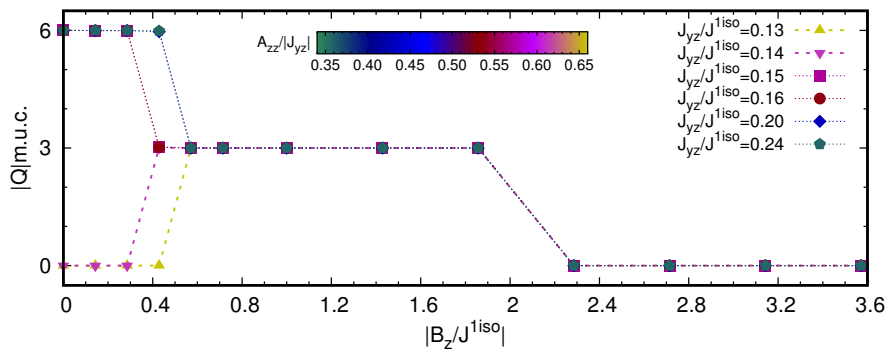}
\caption{Evolution of $|Q|$ per magnetic unit cell ($L_{m.u.c.}=8a_0$), as a function of the magnetic field $B_z$ for different off-diagonal terms, while keeping fixed diagonal exchange parameters and easy-plane anisotropy to values reported in Table~\ref{exchange_param}. Color gradient of the solid points evolves with the $A_{zz}/|J_{yz}|$ ratio. At zero and low $B_z$ field, the system stabilizes a proper screw like the (A)-phase for $J_{yz}/J^{1iso}\leq0.14$, and the topological lattice of the (C)-phase for $J_{yz}/J^{1iso} > 0.14$. At intermediate $B_z$, the system undergoes a topological phase transition to a Bloch-type skyrmion lattice, (E)-phase. At high $B_z$, the transition to the ferromagnetic state, (I)-phase, takes place passing trough the trivial triple-$\bm q$ state, (H)-phase. \label{vario_jyz}}
\end{figure}

In Fig.~\ref{vario_jyz} we report the evolution of the topological charge per magnetic unit cell $|Q|_{m.u.c}$ as a function of increasing $B_z$ and for different values of the off-diagonal $J_{yz}$ exchange coupling term, which is treated here as a measure of the non-coplanarity induced by the two-ion anisotropy. SIA is kept fixed to the NiI$_2$ reference value (Table~\ref{exchange_param}); in fact, from the phase diagram previously discussed (Fig.~\ref{phase_diagram}), we found that the SIA is basically ineffective against the anisotropic symmetric exchange interaction for a wide range of values, that are $-1.50\lesssim A_{zz}/|J_{yz}|\lesssim +2.50$ and $-0.30\lesssim A_{zz}/|J|^{1iso}\lesssim +0.50$. 

For $J_{yz} > 0.14 J^{1iso}$ (reference value is $J_{yz}\simeq 0.20 J^{1iso}$), the topological lattice with $|Q|_{m.u.c}=6$, represented by the (C)-phase (Fig.~\ref{seiQ}a), is the lowest energy spin configuration stabilized by the considered magnetic interactions; in particular, in line with results shown in Fig.~\ref{phase_diagram}a, two sharp topological phase-transitions, signalled by the abrupt change of the total topological charge, are induced under the applied magnetic field: $|Q|_{m.u.c}=6\to3\to0$. The $|Q|_{m.u.c}=3$-state obtained under intermediate $B_z$ is the Bloch-type skyrmion lattice of the (E)-phase; for strong $B_z$, the transition to the trivial ferromagnetic (I)-phase takes place passing through the intermediate (H)-phase, characterized by the topologically trivial triple-{\bf q} state (Figs.~\ref{zeroQ}l and \ref{zeroQ}g, respectively). 

For $J_{yz} \leq 0.14 J^{1iso}$, two topological phase transitions still takes place, but the stable phase at zero and weak magnetic field is now a trivial single-{\bf q} state; $|Q_{m.u.c}|=0\to3\to0$ as a function of $B_z$. In particular, such single-{\bf q} state is found to be a proper-screw spin-spiral. The magnetic phases induced by the applied fields are closely related to those observed when $J_{yz} > 0.14 J^{1iso}$. Therefore, a Bloch skyrmion lattice can be still stabilized by a magnetic field applied on a topologically trivial helical state when the non-collinearity brought in by the two-ion anisotropy is not sufficiently strong to drive a spontaneous high-$Q$ topological lattice.

\section{Discussion} 

In the previous section we have described the rich anisotropy-field phase diagram (Fig.~\ref{phase_diagram}) obtained by tuning strength and direction of the single-ion anisotropy (from easy-plane to easy-axis) as well as upon increasing magnetic field, while keeping fixed the strength of both the isotropic and anisotropic symmetric exchange interactions describing the short-range magnetic interactions of the spin-spin Hamiltonian (\ref{hamiltonian1}) for a centrosymmetric triangular lattice. We found that, with these magnetic interactions reported in Table~\ref{exchange_param}, the spontaneous (C)-phase consisting of a hexagonal lattice of vortices ($m$=+1) and anti-bi-vortices ($m$=-2) carrying a total topological charge of six per magnetic unit cell, 
is a persistent thermodynamically-stable phase in a wide portion of the SIA-field phase diagram; it is in fact, to a wide extent, independent on the easy-plane or easy-axis character of the single-ion anisotropy. Our analysis also unveils different field-induced topological phase transitions that can be realized for different values of the SIA coupling constant. 

\begin{figure}[b]
\centering
\includegraphics[width=14cm]{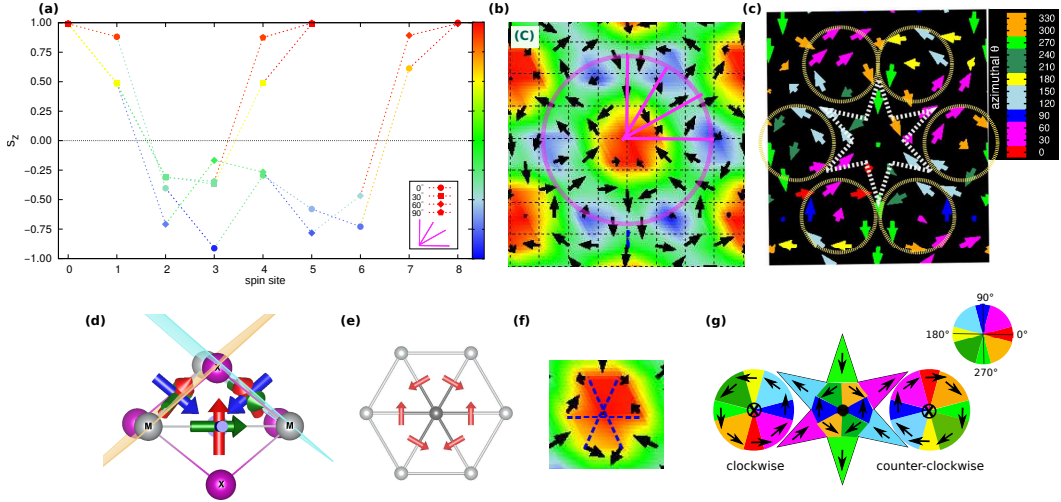}
\caption{Profile of the out-of-plane component of spins and in-plane spin texture of the (C)-phase. ({\bf a}) $s_z$ component of spins for each magnetic site from the MC simulations, when moving from the upward ($s_z=+1$) core (spin site n. 0) of the topological object (A2V) along the radial directions ($0^{\circ},~30^{\circ},~60^{\circ}~\textrm{and}~90^{\circ}$, purple lines) toward the next-neighbor A2V-cores. This defines the $s_z$ profile, which can be regarded as a discretized polarity $p$ associated to the anti-bi-vortex when moving from its core to its finite edge, defined by the distance with respect to the nearest-neighbor downward cores ($s_z=-1$) of the surrounding vortices; the A2V-radius counts $\simeq 4$ spins ({\bf b}). ({\bf c}) In-plane components of spins; arrows are colored with the azimuthal angle $\theta \in [0^{\circ}, 360^{\circ})$, defined by the $s_x$ and $s_y$ components of the spin vector $\bm s$ and highlighting the in-plane spins rotational direction. Schematic representation of the two vortices and anti-bi-vortex together with a color wheel in ({\bf g}) further help visualization of the planar orientation of spins. 
({\bf d}) Lateral view of the local eigenvectors for each $M\textrm{-}X\textrm{-}M\textrm{-}X$ spin-ligand plaquette on the triangular $M$-net to help visualization of the non-coplanarity and non-collinearity in the exchange-tensor principal axes. ({\bf e}) In-plane components of the red eigenvector pointing along the $X\textrm{-}X$ direction for each six nearest-neighbor magnetic  $M\textrm{-}M$ pair. Specific case here concerns the two-ion anisotropy (${\bm J}^{S}$) estimated in monolayer NiI$_2$~\cite{noi} (Table~\ref{exchange_param}). Spins on the nearest-neighbour $M$ sites of the central site orient according to the in-plane projection of the non-coplanar principal axes, as from the zoom on the spin texture of the A2V-core obtained by MC simulations ({\bf f}).
\label{profile_ground}}
\end{figure}

\begin{figure}[]
\centering
\includegraphics[width=14cm]{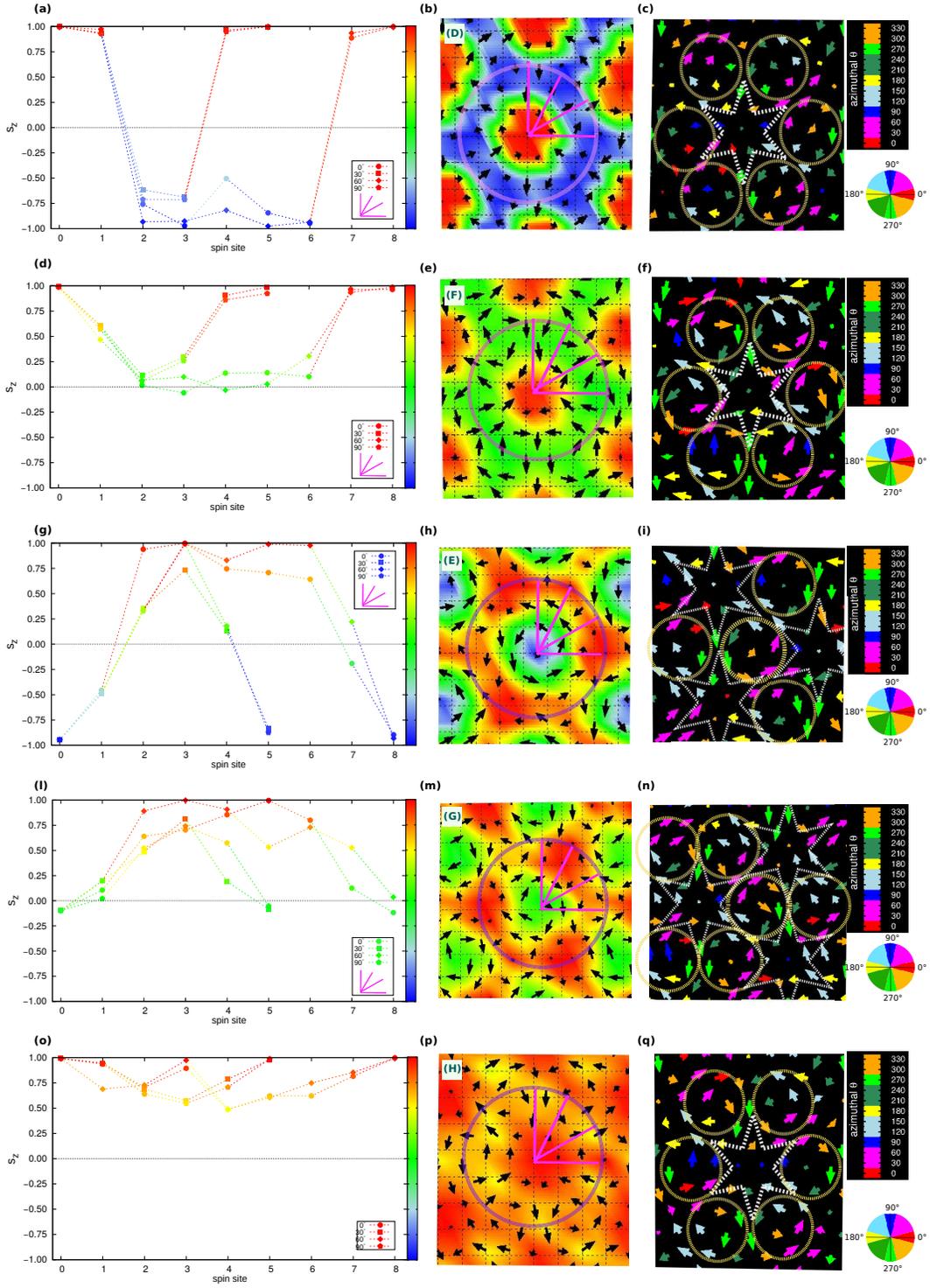}
\caption{Profile of the out-of-plane component of spins ($s_z$) and planar spin texture with associated azimuthal angle defined by the $\{s_x,s_y\}$ components, as in Fig.~\ref{profile_ground}(a,b,c), for the A2SK-type lattice of the (D)-phase ({\bf a,b,c}), the A2M-lattice of the (F)-phase  ({\bf d,e,f}), the Bloch-type SK-lattice of the (E)-phase  ({\bf g,h,i}), the meron lattice of the (G)-phase ({\bf l,m,n}), and the trivial triple-{\bf q} state of the (H)-phase  ({\bf o,p,q}).\label{profile}}
\end{figure}

Nevertheless, it is important to notice that all the different emerging triple-({\bf q}) states, from topological (phases C, D, E, F, G) to trivial (H-phase) spin configurations, exhibit a common planar spin-texture. 
To better appreciate this feature, we show in Fig.~\ref{profile_ground}c and Fig.~\ref{profile}(c,f,i,n,q) the color-gradient plots of the in-plane spin components highlighting the azimuthal angle $\theta \in [0^{\circ}, 360^{\circ})$, the in-plane spins rotational direction, defined by the $s_x$ and $s_y$ components of the spin vector $\bm s$. 

Such specific in-plane orientation of spins can be traced back to the two-ion anisotropy (TIA), which tends to orient the spins along given non-collinear and non-coplanar orientations in space, defining both the helicity and vorticity of the spins pattern and thus behaving as an emergent chiral interaction. Indeed, as discussed in Refs.~\cite{noi,noi_suppl}, the principal magnetic axes per bond of the $\bm J^S$ tensor are not all parallel either to the lattice vectors or with the principal axes of neighbouring bonds, as shown in Fig.~\ref{profile_ground}(d); this introduces non-coplanar components in the spin-spin interaction and, thus, frustration in the relative orientation of the spins, which ultimately arises from the non-coplanarity of the spin-ligand $M-X$ plaquettes mediating the exchange interactions in the spin triangular lattice. In particular, the in-plane projection of the principal axis directed along the ligand $X-X$ direction [indicated by red arrows in Fig.~\ref{profile_ground}(d,e)] fixes the in-plane orientation of the nearest-neighbour magnetic moments giving rise to the A2V spin-pattern [Fig.~\ref{profile_ground}(e,f)]; the accommodation of the anti-bi-vortices in the spin lattice, which tend to overlap because of the short-period modulation due to the isotropic magnetic frustration ($J^{3iso}~vs~J^{1iso}$), causes the emergence of the surrounding vortices. 

With the given strong anisotropic symmetric exchange interaction considered in the present case, the two-ion anisotropy dominates over the single-ion anisotropy and external magnetic field: SIA and $B_z$ tune the out-of-plane spins component, modifying the polarity $p$, and thus the final spins scalar chirality and related topology ($Q$) of the spin configuration, while the planar spin texture result always fixed by the TIA. All the triple-({\bf q}) phases can be in fact seen as a transformation of the initial (C)-phase via major modifications of the $s_z$ component of spins; the in-plane directions of spins remaining unchanged.
Moreover, even though single-$\bm q$ states are found to be energetically more stable than triple-$\bm q$ ones when $A_{zz}\gtrsim 2.5|J_{yz}|$ and $A_{zz}\lesssim -1.5|J_{yz}|$ (or also when $J_{yz}\leq 0.14 J^{1iso}$), an applied out-of-plane magnetic field can still induce a transition to a topological skyrmion lattice, whose in-plane spin configuration appears to be determined by the two-ion anisotropy.


\section{Conclusions}
In this work we have theoretically investigated the effects of competing single-ion and two-ion anisotropies in a triangular lattice with strong magnetic frustration, as occurring in monolayers of van der Waals nickel dihalides. By means of Monte Carlo calculations, we analyzed the parameter space spanned by SIA and applied $B_z$ field for a spin-lattice model whose parameters have been estimated for the monolayer of the prototypical semiconducting NiI$_2$. Our analysis has revealed a rich phase diagram comprising different magnetic phases, from topologically trivial single-({\bf q}) and triple-({\bf q}) states to topological triple-({\bf q}) states. 
The strong magnetic frustration arising from competing isotropic nearest-neighbour FM and third nearest-neighbour AFM exchange interactions promote the onset of short-period helimagnetic configurations, whereas the strong two-ion anisotropy combined with the geometrical frustration of the underlying triangular lattice favours the stabilization of triple-$\bm q$ states. At zero magnetic field, these result in a lattice of vortices and anti-(bi)-vortices carrying a total topological charge per magnetic unit cell $|Q|_{m.u.c}=6$ that is robust within a wide range of single-ion anisotropy strength. Such topological lattice can be interpreted as a fractionalized anti-bi-skyrmion lattice where each anti-bi-skyrmion is surrounded by six vortices arising from the overlap with neighbouring anti-bi-skyrmions, causing a fractionalization of the topological charge of individual anti-bi-vortices. Both the single-ion anisotropy and the applied field acts primarily on the out-of-plane component of the spin configurations, hence on the polarity $p$ of the vortical states, modulating both the size of anti-bi-vortices and the fractionalization of their charges. As a general trend, on the one hand, easy-axis anisotropy is found to increase the localization of anti-bi-vortices, thus reducing the ``spilling'' of topological charge to surrounding vortices; on the other hand, a strong easy-plane anisotropy would eventually favour, when combined with applied field, a crossover to meronic lattices, {\it i.e.} lattices formed by topological objects with half-integer polarity. Nevertheless, an out-of-plane magnetic field, sustained by the strong anisotropic exchange, is always found to trigger a topological transition to a Bloch-like skyrmion lattice, when applied either on a single- or a triple-$\bm q$ state. Interestingly, the in-plane components of the spin texture are found to be extremely robust across almost the whole phase space explored in this work, with 
exceptions only for extreme values of SIA and fields. 

In conclusion, the different topological phases forming the SIA-field phase diagram can be understood by a change of their polarity, which is directly tuned by both the single-ion anisotropy and the applied field, whereas their vorticity can be always traced back to the two-ion anisotropy on the triangular lattice, which is ultimately responsible for the promotion of both topological and trivial triple-$\bm q$ states.
Our study thus corroborates the potential role of frustrated anisotropic symmetric exchange in defining the vorticity of non-collinear and non-coplanar spin configurations, leading to various possible topological lattices in 2D semiconducting magnets.

\newpage
\appendix
\section{}
\label{model_spiral}
The single-$\bm q$ proper-screw (ps) - phase (A) - and cycloidal (c) - phase (B) - helices propagating along the $x$ axis can be modeled, respectively, as
\begin{eqnarray}
\bm s^{ps}_i &=& (0,\,\cos \alpha_i,\,\sin \alpha_i) \nonumber\\
\bm s^c_i &=& (\cos \alpha_i,\, \sin\alpha_i \sin\theta,\, \sin\alpha_i\cos\theta)
\end{eqnarray}
where $\alpha_i=q x_i$ and $\theta$ is the polar angle describing the tilting of the spin-rotation plane in the cycloidal configuration. The corresponding mean-field energies evaluated for model Eq. \ref{hamiltonian1} can be generally written as:
\begin{equation}
E_{helix}(q) = E^{iso}(q) + E^{SIA}_{helix}+ E^{TIA}_{helix}(q)
\end{equation}
where 
\begin{equation}
E^{iso}(q)=2 J^{1iso}\cos \frac{q}{2}+(J^{1iso}+2J^{3iso})\cos q +J^{3iso}\cos 2q,
\end{equation}
for all single-$\bm q$ states, while $E^{SIA}_{ps}=A_{zz}/2$ for the proper-screw spiral and $E^{SIA}_c=A_{zz}\cos^2\theta/2$ for the cycloid. The mean-field expression for TIA contribution is less transparent, but one can easily show that the off-diagonal TIA term $J^{S}_{yz}$ only affects the cycloid mean-field energy contributing as $J^S_{yz}\sin2\theta\,(\cos q/2-\cos q)$.
The interplay of both SIA and TIA removes the degeneracy of all coplanar helical states arising from isotropic magnetic frustration. A strong easy-axis SIA ($A_{zz}< 0$) would promote a proper-screw or a vertical ($\theta=0$) cycloidal helix, with spins rotating in the $yz$ or $xz$ plane, respectively. The degeneracy between these two helices is then removed by TIA, being $E^{TIA}_{ps}-E^{TIA}_c =J_\perp\,(\cos q/2+\cos q)$, where $2J_\perp=J^{S}_{xx}-J^{S}_{yy}$. Since $J_\perp<0$ from Table \ref{exchange_param}, the proper-screw helix of phase (A) is ultimately stabilized over the vertical cycloid by the two-ion anisotropy. In the opposite limit of strong easy-plane SIA ($A_{zz}>0$), the in-plane $\theta=\pi/2$ cycloid is energetically more favourable than the proper-screw spiral, while the tilting of the spin-rotation plane with respect to the $xy$ plane is induced by the strong off-diagonal $J^{S}_{yz}$, pointing to TIA as the microscopic origin of the (B)-phase.



\acknowledgments
{This work was supported by the Nanoscience Foundries and Fine
Analysis (NFFA-MIUR Italy) project. P.B and S.P. acknowledge financial support from the Italian Ministry for Research and Education through PRIN-2017 projects
“Tuning and understanding Quantum phases in 2D materials - Quantum 2D” (IT-MIUR Grant No. 2017Z8TS5B) and “TWEET: Towards Ferroelectricity in two dimensions” (IT-MIUR Grant No. 2017YCTB59), respectively.

Authors acknowledge high-performance
computing (HPC) systems operated by CINECA 
(IsC72-2DFmF grant HP10CCD7O5, IsC80-Em2DvdWd grant HP10CIIUWD and IsB19-EVE grant HP10BKBJMI projects)
and computing resources at the Pharmacy Dpt. University of Chieti-Pescara, also thanking Prof. Loriano Storchi for his technical support.
}


\end{document}